\documentstyle[preprint,aps,epsfig]{revtex}
\input epsf.tex
\def\DESepsf(#1 width #2){\epsfxsize=#2 \epsfbox{#1}}
\begin{document}
\pagestyle{empty}                                      
\preprint{
\font\fortssbx=cmssbx10 scaled \magstep2
\hbox to \hsize{
\hbox{
            }
\hfill $
\vtop{
 \hbox{ }}$
}
}
\draft
\vfill
\title{Indications for Factorization and \boldmath{${\rm Re\,}V_{ub} < 0$}
\\ from Rare B Decay Data
}
\vfill
\author{$^{1}$Xiao-Gang He, $^{1,2}$Wei-Shu Hou, and $^3$Kwei-Chou Yang}
\address{
\rm $^1$Department of Physics, National Taiwan University,
Taipei, Taiwan 10764, R.O.C.\\
\rm $^2$ Physics Department, Brookhaven National Laboratory,
Upton, NY 11973, USA\\
and\\
\rm $^3$Institute of Physics, Academia Sinica, Taipei, R.O.C}

%
%
\vfill
\maketitle
\begin{abstract}
Surveying known hadronic rare B decays,
we find that the factorization approximation
can give a coherent account of $K\pi$, $\pi\pi$ and $\rho^0\pi^+$ data
and give predictions for $\omega^0\pi^+$, $\rho\pi$ and $K^*\pi$ modes,
{\it if ${Re\,}V_{ub}$ is taken as negative} (in standard phase convention)
rather than positive.
As further confirmation, we expect 
a lower $\sin2\beta$ value at B Factories 
as compared to current fits,
and $B_s$ mixing close to LEP bounds at SLD and CDF.
\end{abstract}
\pacs{PACS numbers: 
12.15.Hh, 
13.25.Hw, 
13.25.-k  
}

\pagestyle{plain}

The last few years have been quite exciting for
the field of hadronic rare B decays \cite{jima}.
The observation of exclusive
$B \longrightarrow \eta^\prime K^+$, $\eta^\prime K^0$,
$K^+\pi^-$, $K^0\pi^+$, and $K^+\pi^0$ modes
give definite support for $b\to s$ penguins,
while $\omega h^+$ \cite{omegaK} and especially the
newly observed $\rho^0\pi^+$ mode \cite{gao} indicate that
tree level hadronic $b\to u$ transitions do occur.
In contrast, the limits on $\phi K^+$ and
$\pi^+\pi^-$, $\pi^+\pi^0$ modes are rather stringent \cite{jima,frank}.
Faced with the questions raised by these measurements,
together with the fact that
two new B Factories would turn on this year,
there is a sense of urgency for us to
reach better understanding of these modes.

Admittedly, much uncertainty clouds
the theory of hadronic rare B decays.
The effective Lagrangian that describes
b quark decay is better understood,
but the subsequent evolution of the decayed B meson
into specific {\it light} two body hadronic final states
is certainly very complicated,
while our understanding of long distance QCD is limited.
The usual approach is to assume factorization,
then use parameters such as $N_{\rm eff.} \neq N_C \equiv 3$ to
fit and quantify the apparent deviations from this assumption.
The picture is further muddled by the possibility of
rescattering between hadronic final states (FSI).
Attempts have been made \cite{ali} to take most
uncertainties into account and project into the
future on the many effective two body modes,
where the experimental outlook is rather bright.
But, can the navigation chart be simplified?
In this Letter we make such an attempt at
understanding present data.

We find a simple, coherent and therefore
attractive view that can account for
current trends in data,
especially $K\pi$, $\pi\pi$ and $V\pi$
($V = \rho$, $\omega$ and $K^*$) modes:
Naive factorization works without resort to $N_{\rm eff.}$ or FSI,
but {\it only with $\cos\gamma$ negative},
where $\gamma = {\rm arg}(V_{ub}^*)$ in
the standard phase convention \cite{PDG}.
Smaller light quark masses may also help.
Semi-quantitative predictions can be made which
could be tested in the near future.

Current fits \cite{PDG,fit} to the KM matrix elements, however,
seem to favor $\cos\gamma > 0$.
The preference comes largely from 
the limit on $\Delta m_{B_s}/\Delta m_{B_d}$
where the hadronic uncertainty is restricted to
$\xi^2 \equiv f_{B_s}^2 B_{B_s}/f_{B_d}^2 B_{B_d}$,
which is probably the least uncertain.
With the more conservative
$\Delta m_{B_s} > 10.2$ ps$^{-1}$ \cite{PDG} at 95\% C.L., which also 
corresponds to the current best single experiment sensitivity,
some room is allowed for $\cos\gamma < 0$.
But with $\Delta m_{B_s} > 12.4$ ps$^{-1}$ \cite{fit}
from combining LEP, CDF and SLD data, 
one gets $\gamma \simeq 60$--$70^\circ$ with $\sim 10^\circ$ errors, 
and $\cos\gamma$ seems definitely positive.
We note that the 95\% C.L. contour of one of the fits \cite{fit}
has a tail extending towards $\cos\gamma < 0$,
and would extend further if one enlarged the error on $\xi$.
It may be prudent, therefore, 
to allow for the possibility that $\cos\gamma < 0$
might still be the case in Nature.
The current fit result may be implying that
$B_s$ mixing is not far around the corner.
In any case we should keep in mind that $\gamma$ is 
the most challenging unitarity angle to measure at B Factories,
and any handle one may gain should be welcome.

When 1997 data suggested $K^0\pi^+ > K^+\pi^-$,
a method for constraining $\gamma$ was proposed \cite{FM}.
With 1998 data, the $K^+\pi^0$ mode was observed
while the $K^0\pi^+$ rate came down \cite{jima},
and both branching ratios ($Br$) are now similar to $K^+\pi^-$
$\simeq 1.4\times 10^{-5}$.
Although the method of Ref. \cite{FM} is no longer effective,
it was pointed out \cite{DHHP} that 
the 1998 data suggest $\cos\gamma < 0$ \cite{NR}
and prefer small or no FSI phase.
Following this trail, we find that a
negative $\cos\gamma$ could also explain the absence of
the $\pi^+\pi^-$ mode,
the prominence of $\rho^0\pi^+$ over $\omega^0\pi^+$ and $K^{*0}\pi^+$,
as well as predict emerging trends in
$\pi\pi$, $\rho\pi$ and $K^*\pi$ modes.

Let us retrace the main points of Ref. \cite{DHHP}.
We give the average $K\pi$ branching ratios vs. $\gamma$
in Fig. 1(a) for $m_s = 105$ and 200 MeV.
The light quark mass $m_s$ enters through
the penguin $O_6$ operator via relations between
axial current and pseudoscalar density matrix elements.
We see that $K^+\pi^- \simeq K^0\pi^+ \simeq K^+\pi^0$
prefers a larger $m_s$,
and can only be achieved (allowing for some experimental uncertainty)
for $\gamma \sim 90^\circ - 130^\circ$, or $\cos\gamma <0 $.
Although \cite{DHHP} the electroweak penguin (EWP)
plays a crucial role in raising the $K^+\pi^0$ rate,
the change in sign of $\cos\gamma$ was
important in allowing $K^+\pi^-$ to reach above $K^0\pi^+$.

With present fit values for $V_{ub}$, one expects
$\pi^+\pi^0 < \pi^+\pi^- \sim 1\times 10^{-5}$.
Instead, one finds
$\pi^+\pi^- < 0.84\times 10^{-5}$ \cite{frank}
and a weaker limit on $\pi^+\pi^0$
due to a larger event yield.
Compared to the strength of the $K\pi$ modes,
they pose some problem for theory.
Again, the traditional approach is to
resort to $N_{\rm eff.}$ or FSI, or a smaller $\vert V_{ub}\vert$.
We find, rather interestingly, that a
simple flip in sign of $\cos\gamma$
not only explains the smallness of the $\pi^+\pi^-$ mode,
but also allows for $\pi^+\pi^0 > \pi^+\pi^-$,
without need for very small $N_{\rm eff.}$
or large $\pi^+\pi^- \to \pi^0\pi^0$ rescattering \cite{pipi}.
The amplitude for the $\bar B^0\to \pi^+\pi^-$ mode is,
\begin{eqnarray}
{\sqrt{2}} \, {\cal A}_{\pi^+  \pi^-}  = i{G_F} f_\pi F_0\, (m_B^2-m_\pi^2)
\left\{V_{ud}^*V_{ub} \, a_1
     - V_{td}^*V_{tb} [a_4+a_{10} +(a_6+a_8)R_1]\right\},
\label{pi+pi-}
\end{eqnarray}
where $F_0 = F_0^{B\pi}(m_\pi^2)$ is a $B\to \pi$ (BSW) form factor,
$a_i$'s are combinations of Wilson coefficients \cite{ali},
and $R_1={2m_\pi^2}/{(m_b-m_u)(m_u+m_d)}$.
It is clear that tree--penguin ($T$--$P$) interference
for $K\pi$ and $\pi\pi$ modes differ in sign,
because the KM factors
${\rm Re}\, (V_{ts}^* V_{tb}) \cong -A\lambda^2$ and
${\rm Re}\, (V_{td}^* V_{tb}) \cong A\lambda^3(1-\rho)$
have opposite sign.
{\it
This observation is independent of factorization assumption.
}
As a consequence, if $K^+\pi$ rates are enhanced for $\cos\gamma < 0$, 
the $\pi^+\pi^-$ rate gets suppressed.
In contrast, the $\pi^+\pi^0$ mode is mainly $T$
plus small EWP terms,
hence its $\gamma$ dependence is weak.
Analogous to the $K\pi$ case, 
$u$ and $d$ quark masses enter through $R_1$.
We plot $Br$ vs. $\gamma$ for $\pi\pi$ modes in Fig. 1(b)
for $m_d = 2m_u =$ 3 and 6.4 MeV.
These quark masses are at the $m_b$ scale,
and are within the range given by Particle Data Group \cite{PDG}.
It is clear that
$\pi^+\pi^- < \pi^+\pi^0$ is not impossible for $\cos\gamma < 0$
if $m_{u,d}$ are on the lighter side.
In this case, however,
$P$ would become comparable to $T$, 
complicating mixing dependent CP study in $B^0\to \pi^+\pi^-$ channel.
We note that in general the $\pi^0\pi^0$ mode is very small,
which would not be the case if $\pi^+\pi^-$
is suppressed by rescattering into $\pi^0\pi^0$.

The $\rho^0\pi^+$ mode has just been observed at the
sizable rate of $(1.5\pm 0.5\pm 0.4)\, \times 10^{-5}$ \cite{gao},
and is seemingly larger than $\omega^0\pi^+ \sim 1\times 10^{-5}$
as indicated in \cite{omegaK}.
Both are at odds with
the results of Ref. \cite{ali} for $N_C = 3$.
Can changing the sign of $\cos\gamma$ help?
Dropping EWP terms (but not numerically),
the $B^-\to \rho^0\, (\omega^0)\, \pi^-$ amplitude is
\begin{eqnarray}
{\cal A}_{V^0\pi^-} =   
	G_F m_V \varepsilon\cdot p_\pi
\left\{f_\pi A_0 
     \left[ V_{ud}^*V_{ub} a_1 - V_{td}^*V_{tb} (a_4 + a_6 Q_1) \right]
     + f_{V} F_1 
     \left[ V_{ud}^*V_{ub} a_2 \pm V_{td}^*V_{tb} a_4 \right]
\right\}, \nonumber
\end{eqnarray}
where
$Q_1 = -{2m_\pi^2}/{(m_b+m_u)(m_u+m_d)}$
is opposite in sign to $R_1$ of Eq. (1),
$A_0 = A_0^{BV}(m_\pi^2)$ and $F_1 = F_1^{B\pi}(m_{V}^2)$
are BSW form factors \cite{ali}.
The $+/-$ sign for the last term is for $\rho^0/\omega^0$,
and is traced to the $d\bar d$ content (PDG convention) 
of $\rho^0$ and $\omega^0$ when $\pi^+$ comes from the spectator quark
in a $\bar b\to \bar dd\bar d$ transition.
As shown in Fig. 2(a),
it splits $\rho^0\pi^+$ upwards from $\omega^0\pi^+$
for $\cos\gamma <0$.
Because the difference between the two amplitudes is otherwise minute,
this is a test for $\cos\gamma <0$ independent of normalization.

The normalization is still of some concern for $N_C = 3$.
To see how it might come about,
we note that the $a_4+a_6 Q_1$ term fortuitously cancels
to within 10\% for $m_u + m_d = 9.6$ MeV.
But if $m_u + m_d = 4.5$ MeV for example,
then $a_4 + a_6 Q_1 > - a_6 > 0$ which
would push up $\rho^0\pi^+$ and $\omega^0\pi^+$ 
for $\cos\gamma < 0$ (see Fig. 2(a)).
Scaling up $f_\pi A_0^{BV}$ now by $\sim 20$--30\% brings
these rates above $1\times 10^{-5}$.
For higher $m_u + m_d$ values
a larger $f_\pi A_0^{BV}$ value is needed.
The other possibility of scaling up $f_V F_1^{B\pi}$
runs against the (updated \cite{frank}) limit
$\phi^0 K^+ < 0.59 \times 10^{-5}$,
which is proportional to $f_\phi F_1^{BK}$ in amplitude.
This mode is also plotted in Fig. 2(a), and 
a slight reduction of $f_\phi F_1^{BK}$ seems to be needed.
The $\phi^0K^+$ rate is unaffected by $m_{u,d,s}$ since
the $\phi^0$ vector meson cannot come from the spectator quark
in $B^+$ decay.

For $\cos\gamma > 0$ and $N_{\rm eff.} = 3$ ($2$)
one expects \cite{ali} the combined $\rho^\pm\pi^\mp$
(separating $B^0$ from $\bar B^0$ decay requires tagging)
and $\rho^+\pi^0$ rates to be $\sim 7$ ($4$) and
$3$ ($2$) times the $\rho^0\pi^+$ rate, respectively,
which are very sizable.
It is interesting that,
while the $\rho^0\pi$ rates are enhanced for $\cos\gamma < 0$,
the $B\to \rho^+\pi$ rates are suppressed.
Thus,
lower $\rho^+\pi^-/\rho^0\pi^+$ and $\rho^+\pi^0/\rho^0\pi^+$ ratios
would also suggest that $\cos\gamma < 0$ is preferred.
We plot these effects in Fig. 2(b), again for
$m_d = 2m_u =$ 3 and 6.4 MeV.
Note that the $B^0\to \rho^+\pi^-$ mode is
insensitive to $m_{u,d}$.
The combined $Br(B^0\to \rho^\pm\pi^\mp)$ is
still likely to be over $4$ times larger than $\rho^0\pi^+$, 
and since the final state contains only one $\pi^0$,
it should be observed soon [See Note Added.].

Experimental sensitivities in
$\rho\pi$, $K^*\pi$ and $\rho K$ modes are similar.
With the $\rho^0\pi^+$ observation,
a limit on $K^{*0} \pi^+$ is also reported.
The event yields \cite{gao} suggest that
$K^{*0}\pi^+ > \rho^0\pi^+$ is unlikely,
which seems again at odds with factorization results \cite{ali}
for $\cos\gamma > 0$.
While too early to draw a conclusion,
our earlier argument suggests that
$\rho^0\pi^+ > K^{*0}\pi^+$ is possible for $\cos\gamma < 0$,
especially since $K^{*0}\pi^+$ is insensitive to $\gamma$ and
perhaps suppressed by $f_{K^*} F_1^{B\pi}$ like the $\phi K$ mode.
We plot all the $K^*\pi$ modes in Fig. 3(a).
The $\gamma$ dependence is similar to the $K\pi$ modes of Fig. 1(a),
but there is no sensitivity to $m_s$
since $K^*$ is produced by vector currents.
Thus, independent of $m_s$ and normalization,
we predict that
$K^{*+}\pi^- > K^{*+}\pi^0 \sim K^{*0}\pi^+$ [See Note Added.] 
for $\cos\gamma <0$,
while $K^{*0}\pi^0$ is $\sim$ factor of two lower.
In contrast, 
$\gamma \simeq 60^\circ$--$70^\circ$ \cite{fit} would give
$K^{*0}\pi^+ \sim K^{*+}\pi^- > K^{*+}\pi^0 \gtrsim K^{*0}\pi^0$.

The $\rho K$ modes are analogous to $K^*\pi$ but with
vector meson coming from the spectator quark.
The tree contribution is color suppressed,
so the rates are very sensitive to 
the penguin combination of $a_4 + a_6 Q$,
where $Q = -{2m_K^2}/{(m_b+m_q)(m_q+m_s)}$.
For $m_s = 105$ MeV, this term again largely cancels.
Together with smaller form factors,
the $\rho K$ modes are in general
much lower than the $K^*\pi$ modes,
with $\rho^0K^0$ the largest for $\cos\gamma < 0$.
The cancellation between $a_4$ and $a_6$, however,
is less effective for {\it larger} $m_s$, which could
enhance (suppress) the $\rho K^+$ ($\rho K^0$) modes considerably
for $\cos\gamma < 0$, as can be seen from Fig. 3(b).
Thus, they could provide useful tests for $m_s$.
Note that if the prominence of $\rho^0\pi^+$ is
in part due to a larger $A_0^{B\rho}$,
then some of the $\rho K$ modes could be $\sim 0.5\times 10^{-5}$.
However, these modes are too sensitive to $m_s$
for one to make firm predictions.

For the very prominent $\eta^\prime K$ modes,
the $g^*\to g\eta^\prime$ ``anomaly" effect \cite{etapXs} 
that seems to account for semi-inclusive $B\to \eta^\prime + X_s$,
though still controversial,
has to be treated properly.
However, we do not know how to treat the 
possible $\vert\bar s g q \rangle$  Fock component of the $K$ meson.
Since in general penguins dominate,
the rates are not very sensitive to $\gamma$,
but one still has the nice feature that
$\eta^\prime K^+$ could be enhanced by 10--20\% 
over $\eta^\prime K^0$ for $\cos\gamma < 0$.

Direct CP asymmetries ($a_{\rm CP}$) 
can arise via penguin absorptive parts.
The $K\pi$ modes have been discussed elsewhere \cite{DHHP}.
The CP eigenstate $\pi^+\pi^-$ may have $a_{\rm CP} \sim$ 15 (10) \% 
for $\cos\gamma < (>)\ 0$,
opposite in sign to that of $K^{(*)} \pi$ modes,
and measurement requires tagging \cite{DeHe}.
The $a_{\rm CP}$ for $\pi^+\pi^0$ is very small
since strong penguin is absent by isospin symmetry.
The $K^*\pi$ and $\rho\pi$ modes are interesting 
since $T/P$ and $P/T$ are respectively of order 20--30\%.
As shown in Fig. 4, 
$a_{\rm CP}$s for $\cos\gamma < 0$ would be smaller (larger) 
in $K^{*+}\pi$ and $\rho\pi^+$ ($\rho^+\pi$)
compared to $\cos\gamma > 0$ case \cite{ali2},
and would again test our conjecture.
The $a_{\rm CP}$s for $K^{*0}\pi$ are small, 
but like $K^0\pi$ modes
a sizable $a_{\rm CP}$ would signal the presence of FSI phases \cite{DHHP}.
The large $a_{\rm CP}$ in $\rho^0\pi^0$ 
corresponds to a very small rate 
and requires tagging to measure.

We offer some remarks before closing.
First,
as shown in Fig. 2(a), we are still 
unable to account for the $\omega^0 K^+$ rate \cite{omegaK}.
However, at the present level of statistics, and
out of ${\cal O}(10)$ measurements or limits,
having a problem or two is perhaps a virtue.
Second, 
we have not discussed $VV$ modes.
They in general depend on several $B\to V$ form factors,
while their detection would likely come
after prominent $PP$ and $VP$ modes.
There is some indication for the $\phi K^*$ mode \cite{omegaK},
but being pure $b\to s$ penguin, it has little bearing on $\gamma$.
Third,
the electroweak penguins have been numerically included.
They are in general less significant than varying $\cos\gamma$.
Four,
larger $a_2$ (or lower $N_{\rm eff.}$)
can \cite{ali} enhance $h^+\pi^0$ ($h = \pi$, $K$, $\rho$ and $K^*$)
and $\rho^0\pi^+$, $\omega^0\pi^+$ modes.
Five,
although we have kept a range for light quark masses,
we note that for $\cos\gamma < 0$,
lower $m_u$, $m_d$ and $m_s$ values 
lead to interesting results such as
further suppressing (enhancing) the $\pi^+\pi^-$
($\rho\pi^+$ and $\omega^0\pi^+$) mode(s),
but making the $\rho K$ modes difficult to predict.
They also suggest the ordering
$K^+\pi^- > K^0\pi^+ \simeq K^+\pi^0 > K^0 \pi^0$
for the $K\pi$ modes.
Finally,
it is surprising that
factorization seems to account for present data
if one simply changes $\cos\gamma$ from positive to negative,
although the latter change runs against 
fits to KM matrix elements \cite{fit}.
That something as simple as factorization
would work for rare hadronic B decays
should be welcome,
and it is further encouraging that the conjecture can be 
tested as more data unfolds, 
where one can perhaps even contemplate
making a more systematic fit to model parameters
in the near future.
If the $\cos\gamma$ value from such fits continues to be 
at odds with updated CKM fits, we may be in store for 
some exciting physics at the B Factories or elsewhere.
For example, $\sin 2\beta$ would be lower
than the CKM fit prediction and more consistent with $\cos\gamma < 0$,
and $B_s$ mixing would be measured soon at the Tevatron
and/or SLD, or else we may have new physics.

In conclusion,
we find the surprising result
that {\it a simple change in sign for $\cos\gamma$
from current fit values can account for present
rare B decay data within factorization approximation}.
The size of the $K\pi$ modes
and the newly observed $\rho^0\pi^+$ mode,
the absence of $\pi^+\pi^-$ (perhaps below $\pi^+\pi^0$) etc.,
can all be due to having constructive rather than destructive 
tree-penguin interference, or vice versa.
Prominence of $\rho^0\pi^+$ probably implies a larger 
$A_0^{BV}$ form factor,
while absence of $\phi^0 K^+$ suggests a smaller $F_1^{BP}$,
which may also contribute to the absence of $K^{*0}\pi^+$.
Chief predictions for $\cos\gamma < 0$ are:
$\rho^0\pi^+ > \omega^0\pi^+$, $K^{*+}\pi^- > K^{*0}\pi^+$,
reduced but still prominent $\rho^+\pi^-/\rho^0\pi^+$
and $\rho^+\pi^0/\rho^0\pi^+$ ratios,
and $K^+\pi^- > K^0\pi^+$ if $m_s$ is on lighter end.
One expects a lower $\sin2\beta$ value at B Factories
compared to current fit results,
and $B_s$ mixing close to present LEP bounds..

This work is supported in part by
grants NSC 88-2112-M-002-033, NSC 88-2112-M-002-041 
and NSC 88-2112-M-001-006 of the Republic of China.
We thank J. Alexander, H.Y. Cheng, Y.S. Gao, W. Marciano, 
A. Soni, J. Smith, W. Sun, F. W\" urthwein and
L. Wolfenstein for discussions.
WSH thanks the Theory Group of Brookhaven National Lab for partial support,
and K. Berkelman, G. Brandenburg and E. Thorndike for hospitality 
during frequent visits to the CLEO Collaboration at Cornell University.

\vskip 1.5cm

\noindent {\bf Note Added.}

After this work was posted,
CLEO announced \cite{SGW} the measurement of 
$Br(B\to \rho^\pm\pi^\mp)=(3.5^{+1.1}_{-1.0}\pm 0.5)\times 10^{-5}$
and 
$Br(B\to K^{*+}\pi^-) = (2.2^{+0.8+0.4}_{-0.6-0.5})\times 10^{-5}$,
which further confirm our conjecture that $\cos\gamma < 0$.
The ratio $\rho^\pm\pi^\mp/\rho^0\pi^+ \simeq 2.3$ turns out to be
less than 4 which we had advocated.
From hindsight, since ${\cal A}(B^0\to \rho^+\pi^-)\propto F_1^{B\pi}$, 
this can be attributed to our observation that $A_0^{BV}$ is
enhanced to account for $\rho^0\pi^+$ rate,
while $F_1^{B\pi}$ is suppressed as indicated by
$\pi^+\pi^-$ and $\phi K^+$ nonobservation.

\begin{figure}[htb]
\vspace{0.5cm}
\centerline{ \DESepsf(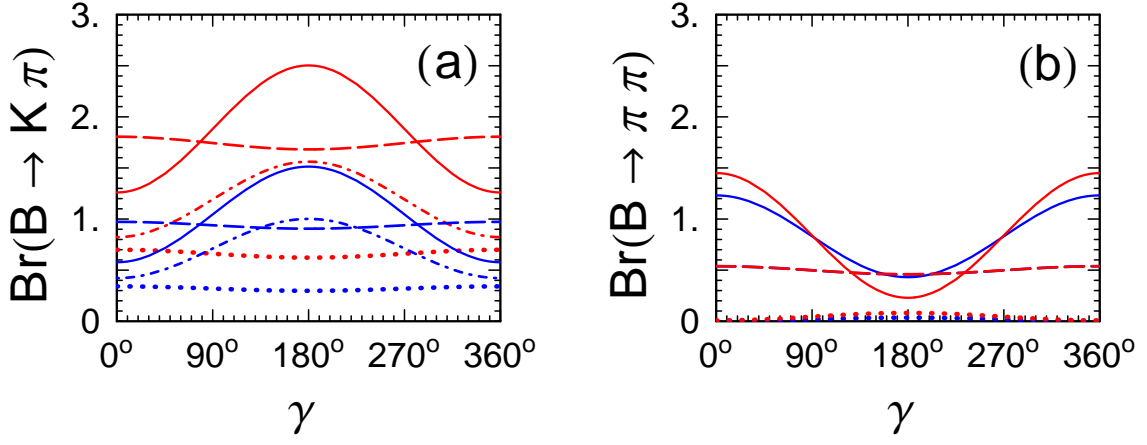 width 15 cm) }
\smallskip
\caption {(a) Solid, dash, dotdash and dots for
$B\to K^+\pi^-$, $K^0\pi^+$, $K^+\pi^0$ and $K^0\pi^0$,
for $m_s = $ 105 (upper curves) and 200 MeV.
(b) Solid, dash and dots for
$B\to \pi^+\pi^-$, $\pi^+\pi^0$ and $\pi^0\pi^0$
for $m_d = 2 m_u= $ 3 and 6.4 MeV,
where the lower (upper) curve at $\gamma = 180^\circ$
for $\pi^+\pi^-$ ($\pi^0\pi^0$) is for lower $m_{u,d}$.
In all figures $Br$s are in units of $10^{-5}$,
and $\vert V_{ub}/V_{cb}\vert = 0.08$. }
\label{PP}
\end{figure}

\begin{figure}[htb]
\vspace{0.5cm}
\centerline{ \DESepsf(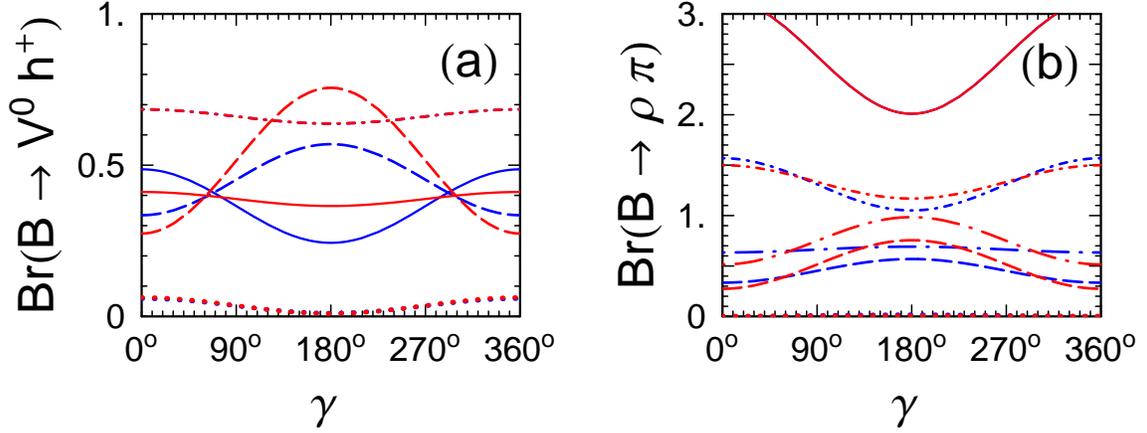 width 15 cm) }
\smallskip
\caption {
For $m_d = 2m_u= $ 3 and 6.4 MeV,
(a) solid, dash, dotdash and dots for
$\omega^0\pi^+$, $\rho^0\pi^+$, $\phi^0 K^+$ and $\omega^0 K^+$;
(b) solid, short-dotdash, long-dotdash, dash and dots for
$B\to \rho^+ \pi^-$, $\rho^+\pi^0$, $\rho^-\pi^+$,
$\rho^0 \pi^+$ and $\rho^0\pi^0$.
The upper curves at $\gamma = 180^\circ$
are for lower $m_{u,d}$.
}
\label{VP}
\end{figure}


\begin{figure}[htb]
\vspace{0.5cm}
\centerline{ \DESepsf(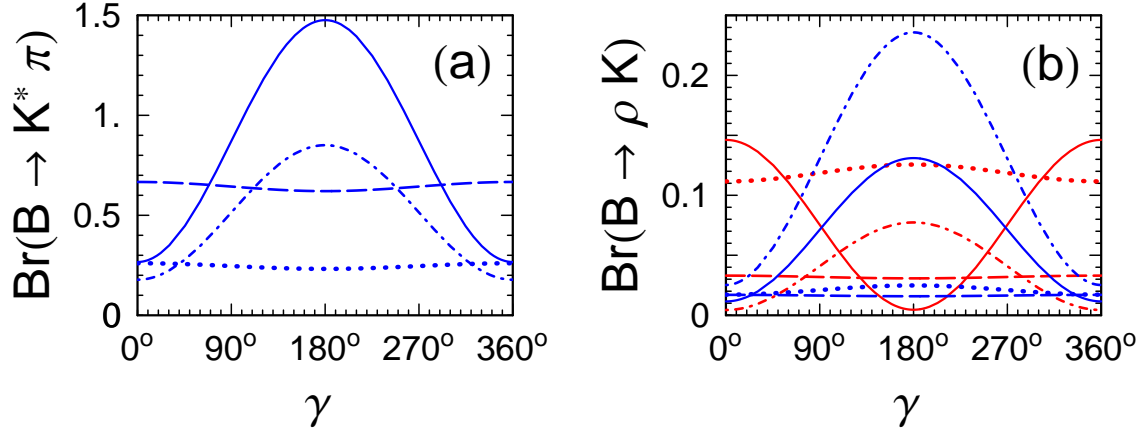 width 15 cm)}
\smallskip
\caption {
(a) Solid, dash, dotdash and dots for
$B\to K^{*+}\pi^- $, $K^{*0}\pi^+$, $K^{*+}\pi^0$ and $K^{*0}\pi^0 $,
which are insensitive to $m_s$.
(b) Solid, dash, dotdash and dots for $\rho^- K^{+}$,
$\rho^+ K^{0}$, $\rho^0  K^{+}$ and $\rho^0 K^{0}$,
for $m_s =$ 105 and 200 MeV.
The upper (lower) curves for $\rho K^0$ ($\rho K^+$)
at $\gamma = 180^\circ$ are for lower $m_s$.
} \label{rhoK}
\end{figure}

\begin{figure}[htb]
\vspace{0.5cm}
\centerline{ \DESepsf(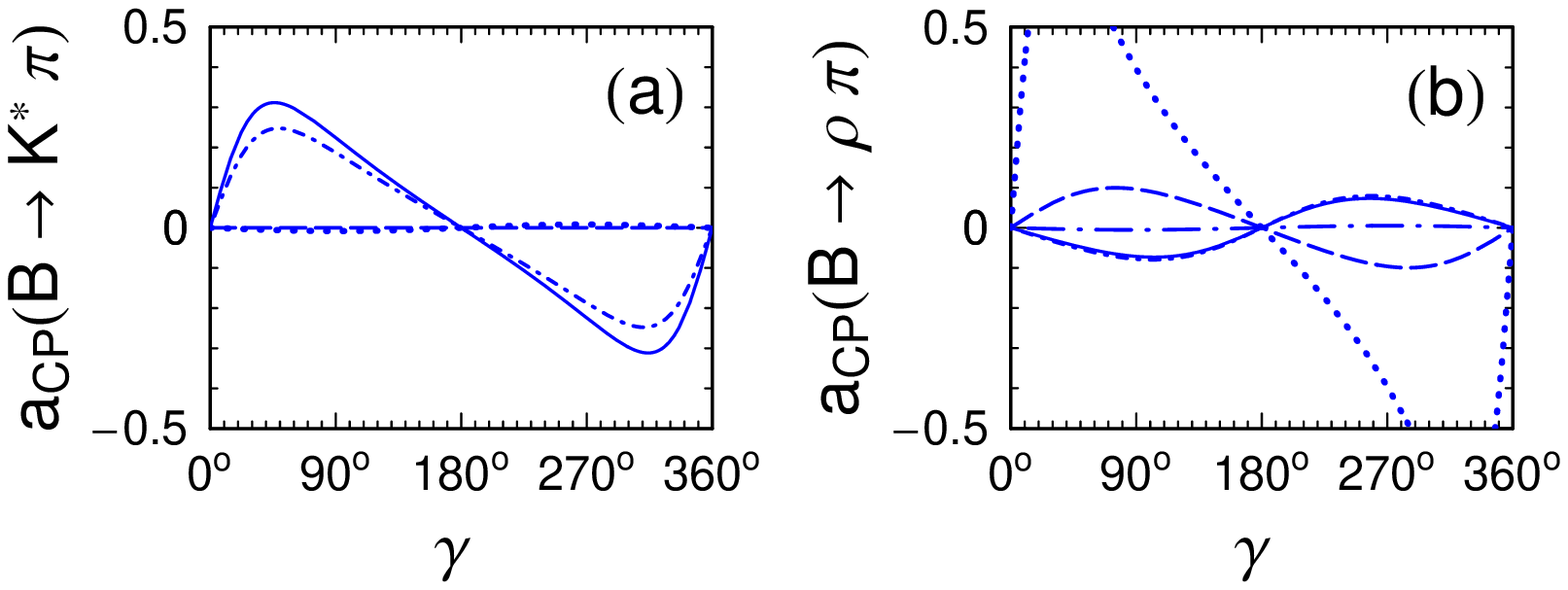 width 15 cm) }
\smallskip
\caption {
Direct CP violating asymmetries vs. $\gamma$ for
(a) $K^*\pi$ and
(b) $\rho\pi$ modes (for $m_d = 2 m_u= $ 6.4 MeV),
with same notation as in Figs. 3(a) and 2(b),
respectively,
and with $q^2 = m_b^2/2$ for penguin absorptive parts.
}
\label{VPaCP}
\end{figure}


\begin{references}
\bibitem{pipi} FSI rescattering could change the relative size of 
$B^0\to \pi^+\pi^-$ and $\pi^0\pi^0$.
The analysis is similar to the one carried out in Ref.\cite{DHHP}.
\end{references}
\end{document}